\documentclass[acmsmall]{acmart}
\usepackage{tabularx}
\usepackage{colortbl}
\usepackage{multirow}
\AtBeginDocument{%
  \providecommand\BibTeX{{%
    \normalfont B\kern-0.5em{\scshape i\kern-0.25em b}\kern-0.8em\TeX}}}

\setcopyright{acmlicensed}
\copyrightyear{2024}
\acmYear{2024}
\acmDOI{XXXXXXX.XXXXXXX}

\acmConference[CSCW '24]{}{November 09--13,
  2024}{San José, Costa Rica}
\acmISBN{978-1-4503-XXXX-X/18/06}

\begin{document}

\title{The Unwanted Dissemination of Science: The Usage of Academic Articles as Ammunition in Contested Discursive Arenas on Twitter}

\author{Richard Zhang}
\affiliation{%
  \institution{Northwestern University}
  \country{USA}}

\author{Em\H{o}ke-\'Agnes Horv\'at}
\affiliation{%
  \institution{Northwestern University}
  \country{USA}}
  

\begin{abstract}
  Twitter is a common site of offensive language. Prior literature has shown that the emotional content of tweets can heavily impact their diffusion when discussing political topics. We extend prior work to look at offensive tweets that link to academic articles. Using a mixed methods approach, we identify three findings: firstly, offensive language is common in tweets that refer to academic articles, and vary widely by subject matter. Secondly, discourse analysis reveals that offensive tweets commonly use academic articles to promote or attack political ideologies. Lastly, we show that offensive tweets reach a smaller audience than their non-offensive counterparts. Our analysis of these offensive tweets reveal how academic articles are being shared on Twitter not for the sake of disseminating new knowledge, but rather to as argumentative tools in controversial and combative discourses.
\end{abstract}



\maketitle

\section{Introduction}
The dissemination of information on online social media is pertinent within variegated contexts, such as political, health and crisis messaging. Twitter, known as ``X'' as of July 2023, is a popular social media site that allows its users to emit short-form messages to their network of followers, which were formerly known as tweets. Within politics, Twitter is used by politicians to advocate for their political agendas and campaigns. For health news, diffusion of health policies and recommendations inform the public in a timely manner \cite{kullar-ea-2020}. Warnings of environmental hazards and emergencies were often transmitted and retransmitted on Twitter \cite{sutton}. Academics are active on Twitter to increase the visibility of their work and networks. Though some of these uses of Twitter can be characterized as healthy, productive, or innocuous, Twitter is also a common site of harmful behavior such as aggression, bullying, and misinformation.

This paper contextualizes how scholars' academic works are used on what formerly was Twitter, where a user may link or cite a paper for a variety of reasons. Academics and researchers often promoted their articles on Twitter \cite{zakhlebin-ea-2020}. A greater dissemination of research articles had been shown to favorably impact scholars' Twitter following and citation counts \cite{ortege-2016, luc-et-al}. These are examples of the productive forces of science; new papers and new knowledge are disseminated to a wider range of people, and hence their authors receive recognition from it. This paper is focused on the more inconspicuous and perhaps more unwanted side of dissemination of science. In particular, we aim to identify how science can be used as fuel in discursive social media arenas. Furthermore, we aim to see how science is being utilized on Twitter as unproductive tools of harassment and argumentation. It has been studied that science plays a role in contested discursive arenas and in small and specific sites \cite{buchanan, anderson-huntington-2017, Hmielowski-ea-2013}, but broad overviews of how science is weaponized have not been conducted generally and broadly across different discursive arenas on social media. To find and analyze how science is utilized negatively in discursive arenas, we identified how science was used in conjunction with offensive language on Twitter. 

Offensive language is defined as hurtful, derogatory, or obscene utterances, and includes insults, profanity, and general abuse \cite{Wiegand, zampieri-ea-2019}. The study of offensive language is particularly salient to the discourse surrounding the academic community and science in general. One pertinent aspect of this is the skepticism and politicization that are attached to academia. Politicization of science and distrust of science are well documented \cite{gauchat-2012}. This politicization of science can elicit offensive language in discussions of science on Twitter. Discussions of science and academia on Twitter have underlying political leanings, which can encourage the use of abusive language \cite{anderson-huntington-2017, mosleh-ea-2021}. Topics such as global warming invite partisan media attacks and skepticism \cite{Hmielowski-ea-2013}. Certain academic disciplines such as political science and sociology are subject to partisan lines by nature. What are offensive tweets talking about, and how do they fall under broader discursive subjects? These results can have broader implications; Gauchat \cite{gauchat-2012}, for instance, suggests that higher visibility and science education leads to distrust, rather than trust, of science in conservative crowds. Furthermore, offensive language in academic contexts can detract from the credibility of researchers' findings \cite{barnes-2018, konig-2019}. By using a mixed-methods approach with topic modelling and critical discourse analysis (CDA), we identify the topics of tweets containing offensive language that reference academic articles, as well as the broader discourses that they fall under. Lastly, we want to distinguish whether academic articles are utilized to correct misconceptions about scientific subjects or used in more malicious ways, and ask if the practice of aggressively citing academic articles is employed by users from one or many ideological attitudes. In essence, we ask:

\textbf{RQ1}: How do offensive tweets citing academic articles fall within broader discursive and ideological boundaries? 

To analyze the effects of offensive language in these tweets, we test the effects of offensive language in the dissemination of these tweets. The goal of scholarly work is ostensibly to produce new knowledge, but previous literature on tweets' virality suggests that tweets with more  more pronounced sentiment disseminated more widely \cite{zafra-ea-2021, tsugawa-ea-2017, tsugawa-ea-2015, stieglitz-ea-2013, antypas-ea-2023, pivecka-ea-2022}. The case that offensive language positively impacts the spread of academic articles would prove unfavorable --- possibly hurting reputations of both academics and academia, and shifting the motive of sharing articles away from sharing knowledge to less auspicious reasons. Thus, is the visibility of tweets citing academic literature in conjunction with negative language higher? Do offensive tweets that reference academic articles increase the virality of content, similar to the political, negatively valenced tweets? A study of their virality can provide meaningful insights to whether or not these tweets reach larger audiences, and whether they reach them quicker than non-inflammatory tweets citing academic articles. We return to the idea of comparing productive dissemination of science with unwanted ones; finding that offensively languaged tweets spread wider or faster than their neutral or postitive counterparts may reveal a dangerous paradigm of how science is cited and used on social media. In summary:

\textbf{RQ2}: Does offensive language in tweets that reference academic papers increase or decrease their virality?

This paper makes two main contributions. First, we use a mixed-methods approach that shows for the first time in the context of science dissemination that academic articles are being both weaponized and criticized with offensive language on Twitter. Secondly, we provide empirical grounds that show the nature of where these tweets are found, what they are discussing, and the nature of their spread.

\section{Related Work}

\subsection{Tweet Virality}

Initial work on tweet virality characterized virality as the number of retweets and concluded that tweets containing emotionally charged terms and emoticons affected retweet volume and likelihood, as well as the presence of hashtags, url, and usernames \cite{naveed-ea-2011, suh-ea-2010}. Hong and Davidson \cite{hong-davidson-2010} utilized topics generated from topic modeling, a natural language processing technique that algorithmically generates topics from a corpus of documents, as significant factors affecting retweet likelihood. More concrete measures of virality have been developed and utilized since then. Steiglitz \& Duan-Xuan \cite{stieglitz-ea-2013} measured virality of twitter messages related to German state parliament elections through retweet volume and retweet speed. They define retweet speed as the difference in time between the original tweet and its first retweet \cite{stieglitz-ea-2013}. Tsugawa et al. \cite{tsugawa-ea-2015} extended their measure of retweet speed to the time between an original tweet and its \textit{n}th retweet.

 These papers analyzed sentiment in terms of the valence-arousal space, which has the valence dimension (negative to positive) and the arousal (calm to negative) dimension \cite{posner-2005}. Twsugawa \cite{tsugawa-ea-2015} and Steiglitz \& Duan-Xuan \cite{stieglitz-ea-2013} indicate that any level of nonneutral (e.g. positive and negative) sentiments increase likelihood for retweet volume and retweet speed. Pivecka et al. \cite{pivecka-ea-2022} and Zafra et al. \cite{zafra-ea-2021} borrow from previous techniques within specific political contexts; Pivecka et al. \cite{pivecka-ea-2022} showed that for tweets from Austrian politicians, tweets with high arousal had higher retweet volumes than low arousal, and that tweets with negative valence were associated with decreased retweet volume while tweets with positive valence increased retweet volume. Zafra et al. \cite{zafra-ea-2021} showed an increased retweet volume for tweets with more positive words and a decreased retweet volume for tweets containing more negative words in their dataset of tweets that referenced the Catalan independence referendum in 2017. These results do not show a consistent direction of bias of sentiment on tweet virality and are expected to be closely tied to the content of their datasets. The domain of the dataset is significant; for instance, Zafra et al. \cite{zafra-ea-2021} hypothesized that negative tweets would garner more retweets due to the Spanish unpopularity of the Catalan independence movement. These trends have not been analyzed in regards to academic scholarship, and how academic articles are spread.

\subsection{Relationships Between Virality, Twitter, Science, and Offensive Language}

Virality has substantial bearing on the influence and exposure for academics and their articles. For academic researchers, the dissemination of their work on Twitter can increase their follower count and increase their citation count by way of an increased follower count \cite{ortege-2016}, which becomes salient considering the increasing vitality of publication metrics to a researcher’s success over the last century \cite{fire-guestrin-2019}. In a study of the effects of articles published by a social media network of cardiothoracic surgery news, authors of the referenced papers gained substantially increased citation counts over a one year period and larger twitter followings \cite{luc-et-al}. For researchers, the effects of having a single article or tweet go viral substantially affects their short and long term visibility by gaining increased followers compared to researchers who have not experienced virality \cite{hasan-ea-2022}. 

Despite researchers' recognition of how Twitter can be helpful in developing academic networks, scientists who have left Twitter entirely have cited reasons of increased hostility and right-wing trolls for leaving \cite{nature}. One reason may be that academic disciplines and their related articles can be controversial on social media sites due to the politicization of science. The politicization of science has increased over the past few decades; Gauchat \cite{gauchat-2012} analyzes how conservatives’ distrust of science has increased from 1974 to 2010, and that increased science education and visibility predicts increasing distrust in conservative networks. One of the most visible topics that is subject to politicization is climate change; over the last several decades debates on climate change have become increasingly drawn along political lines \cite{chinn-ea-2020, mccright-dunlap-2011}. This politicization of science can also inform how and when offensive language is used. Incivility, defined as an attack on a person’s character, was found to be used in association with right-leaning political topics in a discourse analysis of Twitter discussions of climate change \cite {anderson-huntington-2017}. With the politicization and controversy of science, analysis of how offensive language emerges in arguments over “correcting” political news can also inform us of the nature of offensive language that occurs in tweets referencing academic articles. In Mosleh et al. \cite{mosleh-ea-2021}, the analysis of tweet chains debating political news were found to increase in their toxicity as debates became heated.

Computational approaches to several facets of offensive language have been researched, such as the development of natural language processing models that perform automatic detection of offensive language on social media \cite{chen-ea-2012, zampieri-ea-2019, tweetnlp} and closely related automatic detection of hate speech \cite{chen-ea-2012, fortuna-ea-2018, Wiegand, zampieri-ea-2019} and cyberbullying \cite{chatzakou-ea-2017}. Silva et al. \cite{silva-ea-2016} proposes a categorization of hate speech on Twitter and social media website Whisper through a lexical comparison with terminology from Hatebase and the FBI’s legal language of hate crimes, forming categories such as “race,” “ethnicity,” “sexual orientation,” and “gender.” Their lexical analysis of tweets reveal that behavior and race are the predominant targets of hate speech \cite{silva-ea-2016}. Conversely, Evkoski et al. \cite{evkoski-ea-2021} seeks the source of hate speech in Slovenian tweets, which reveals that the majority of offensive tweets are emitted by one singular community, and is dominated by political and ideological slants.

\subsection{Critical Discourse Analysis and Topic Modeling}

The most pervasive view of Tweet virality is that virality emerges from a message’s textual content \cite{guerini-2021a, guerini-2021b}. One way scholars have categorized tweets to understand their content is through topic modeling. Topic modeling is a natural language technique to cluster documents, such as tweets, and generate the most representative words for each cluster. The most common method of topic modeling is Latent Dirichlet Allocation (LDA) \cite{blei-ea-2003}. LDA, though commonly used to generate topics on Twitter datasets, assumes that the documents are much longer than the 280 character maximum of a tweet. Various short text topic modeling (STTM) algorithms have been developed for datasets such as Tweets; Xiaobao et al. \cite{nqtm} proposes a neural network based model that optimizes for coherence score and topic uniqueness that outperforms LDA. They utilize a short-text encoding, a novel embedding technique, and negative sampling to achieve greater topic diversity and topic coherence \cite{nqtm}. The implication of this is that tweets belong to fewer topics, and identifying citizenship of a document to a topic can help topics’ interpretations.

Critical discourse analysis (CDA) investigates dynamics of ideology and social problems through analyzing discourse \cite{johnson-mclean-2020}. CDA is relevant for our problem in two ways: firstly, it allows us to contextualize topics generated from topic modeling, which can be incoherent and difficult to explain, and identify their overarching discourses. Secondly, a document-focused discourse analysis colors the language of specific tweets to reveal  relationships between tweet virality, offensive language, and the politicization of science. Jacobs \cite{jacobs-et-al} notes that topic models can act as decompositions for broader “subjects” or “themes,” and the topics act as a “collection of patterns of language use representing those themes.” In its reverse, we can use CDA to group visibly incoherent topics into their broader, more explainable discourses, as suggested by Aranda et al. \cite{aranda-ea-2021}. Performing document-focused CDA over a topic model appropriately fits the assumption that virality of a document emerges from the content of its text; understanding the context of a topic and its broader ideologies reveals qualities of the tweets that belong to it.

\section{Data}
We obtained a dataset from Altmetric LLC that tracks mentions of over 24 million research articles on a range of publicly available internet sources that include news outlets, blogs, Twitter, Facebook, Reddit, and Wikipedia \cite{Altmetric}. Our sample contains tweets mentioning the 9650 most mentioned articles contains the 9,650 most mentioned collected from from their api between the time span of January 1, 2016 to October 8, 2018. A mention explicitly indicates that a URL link has been made to the respective article in the body of the text of the media post in question, such as a tweet, Facebook post, or blog entry. The dataset does not capture posts that respond to those captured within the Altmetric data.. The resulting dataset contains 3,140,493 tweets, including retweets, made by 748,283 unique users. Each row of the dataset contains the original text of the tweet, the DOI being linked in the tweet, a unique post-id, an anonymised identifier for the Twitter user, and whether or not the tweet is a retweet.

We filter the data with the intent of both generating topics as well as evaluating sentiment and offensive language. The content of the linked articles, such as their title, were removed from the texts such that subsequent analyses focuses only on user-generated text. An initial survey showed that a significant number of tweets only included the title; these tweets were removed. The final dataset contains 2,906,058 tweets made by 748,104 users. Of these tweets, 760,799  are original tweets, and 2,145,259 are retweets. Since our dataset does not indicate the source tweet of a retweeted tweet, we created a retweet dataset by grouping all tweets by their text and DOI and selecting only unique pairs of text and DOI for each unique tweet. By this, any retweets with matching text and DOI must necessarily be a retweet of an original tweet of a text and DOI pair. Functionally, this also means eliminating any tweets that have repeat text and repeat DOI, further filtering our 760,799 unique tweets to 295,078 unique tweets that have 0 or more retweets. 

We assigned each tweet an academic subject with the use of OpenAlex. OpenAlex is a catalog of academic entities and their relations; it archives “works” (such as journal articles, books, and theses), their authors, their subject matters, sources, institutions, and publishers within a relational graph structure. OpenAlex organizes their subjects with a “level” system; highest level subjects, such as Physics or Medicine, are at “level 0,” and deeper levels contain more specificity. They label works’ subjects with a classifier trained on Microsoft’s Academic Graph, a heterogeneous graph of academic articles that utilizes automatic concept tagging \cite{openalex}. We queried each DOI in our dataset for their level 0 concepts, and assigned each tweet with the associated concept of the DOI being referenced. There are a total of 19 level 0 concepts that we organize them by.

\section{Methodology}
\subsection{Offensive Language and Sentiment Classification}
To find which tweets are offensive, and to analyze sentiment and offensive language as factors affecting virality, we used offensive language and sentiment classification. We use the TweetEval model within the Python package TweetNLP to perform both offensive language classification and sentiment analysis. It classifies sentiment analysis by its valence (negative to positive). We output valence as a binary variable for positive, neutral, and negative. TweetEval is a RoBerta-based self-supervised language model that has been trained on SemEval 2019’s offensive language and SemEval 2017’s sentiment challenges as a baseline and was later pre-trained on a corpus of 60M tweets \cite{tweetnlp}. For these tasks, we filtered the dataset to the 760,799 original tweets in our dataset.

We compile the total number of tweets per discipline, offensive tweets, and the percentage of tweets that are offensive. We then test for correlations between discipline and rates of offensive language with a Kruskal-Wallis rank sum test.

\subsection{Topic Modeling and Topic Interpretation}
We use Xiaobao et al.'s \cite{nqtm} implementation of Negative sampling and Quantization Topic Model (NQTM) over the tweets that were classified as offensive in the data annotation step. To optimize the number of topics, we perform a grid search over number of topics $K=[10,25,50,100]$ and select the topic model with the highest coherence score $c_v$ with a reasonable topic uniqueness.

We then perform CDA over the topics to generate broader discourses into which we can group the topics. Afterwards, we categorize each topic with respect to their linkages to different contexts, as suggested by Aranda et al. \cite{aranda-ea-2021}. We perform CDA on these topics by qualitatively analyzing the terms associated with each topic and their most representative documents. We select the 5 most representative documents for each topic $i \in [1, K]$, where K is the number of topics selected in our topic model. This can be written as $argmax(\mathbf{\Theta}_{:,i})[-5:]$, where $\Theta$ is the document-topic distribution matrix of size $D \times K$, and $argmax(\mathbf{\Theta}_{:,i})$ ranks the most representative documents (e.g., a document with the score of ``1'' would be distributed solely to topic $i$, and thus be most representative). To contextualize topics, we also compile the most frequent topics for each discipline $Disc$. For this step, we assign each offensive tweet $i\in Disc$ the topic with the highest value in their topic distribution, $argmax\Theta_{i,:}$, where $Disc$ is the academic discipline their referenced article belongs to. We use these contexts to inform our CDA to generate \textit{topic categories}. By analyzing topics' most representative texts, and integrating them within broader themes of offensive language, politicization, and their relationships to academic disciplines, we can formulate broader topic categories that can contextualize our more granular topics generated by topic modeling.

\subsection{Tweet Virality}
We use regression analysis to determine the virality of all the tweets in our dataset. We test two measures of virality: retweet frequency and the time between the original and the 25th retweet. For our first regression analyses, we consider the total volume of retweets for every tweet in our dataset $numRTs$ as the dependent variable. Our second set of regression analyses consideres the time between the original tweet and the 25th tweet $timeRT25$. For both models, we consider the following variables:
\begin{itemize}
    \item Negative - Binary variable representing a negative valence assignment from our sentiment classification
    \item Positive - Binary variable representing a positive valence assignment
    \item Followers - Number of Twitter followers the original user had at the time of the tweet
    \item Offensive - Binary variable representing offensive classification
    \item Hash - Binary variable representing presence of one or more hashtags
\end{itemize}
We consider all the above variables following prior literature that has shown that valence (negative, positive), followers, and hashtags are significant factors contributing to the virality of a tweet \cite{zafra-ea-2021, tsugawa-ea-2017, tsugawa-ea-2015, stieglitz-ea-2013, antypas-ea-2023, pivecka-ea-2022}, in addition to the offensive variable.

We use a negative binomial generalized linear model for the analyses for these regressions on our retweet volume $numRTs$ as the mean and standard deviation for $numRTs$ suggests over-dispersion ($\mu=3.81, \sigma=21.83$), as well as for $timeRT25$ ($\mu=7327, \sigma=51091$). Their regression models are the following:
\begin{equation}
    log(numRT)=\beta_0+\beta_1negative+\beta_2positive+\beta_3followers+\beta_4offensive+\beta_5hash \end{equation}
    \begin{equation}
            numRT=e^{\beta_0}\times e^{\beta_1negative}\times e^{beta_2positive} \times e^{\beta_3followers} \times e^{\beta_4offensive} \times e^{\beta_5hash}
\end{equation}
\begin{equation}
    log(timeRT25)=\beta_0+\beta_1negative+\beta_2positive+\beta_3followers+\beta_4offensive+\beta_5hash \end{equation}
    \begin{equation}
            timeRT25=e^{\beta_0}\times e^{\beta_1negative}\times e^{beta_2positive} \times e^{\beta_3followers} \times e^{\beta_4offensive} \times e^{\beta_5hash}
\end{equation}

We perform further analysis by decomposing the \textit{offensive} factor for each tweet that is classified as offensive. For this, we assign each offensive tweet $i$ the topic with the highest value in their topic distribution $\Theta$, $argmax\Theta_{i,:}$. Our second decomposition assigns each offensive tweet with a variable representing \textit{topic categories} generated from our CDA. We evaluate each model's fit with their $R^2$ values. This can be modeled as the following, where $\text{category}_i$ is a placeholder for the topic categories we will discover.
\begin{equation}
    log(numRT)=\beta_0+\beta_1negative+\beta_2positive+\beta_3followers+\beta_4hash + \beta_5category_1 + \dots + \beta_jcategory_{j+4}\end{equation}
    \begin{equation}
            numRT=e^{\beta_0}\times e^{\beta_1negative}\times e^{beta_2positive} \times e^{\beta_3followers} \times e^{\beta_4hash} \times e^{\beta_5category_1} \times \dots \times e{\beta_jcategory_{j+4}}
\end{equation}
\begin{equation}
    log(timeRT25)=\beta_0+\beta_1negative+\beta_2positive+\beta_3followers+\beta_4hash + \beta_5category_1 + \dots + \beta_jcategory_{j+4}\end{equation}
    \begin{equation}
            timeRT25=e^{\beta_0}\times e^{\beta_1negative}\times e^{beta_2positive} \times e^{\beta_3followers} \times e^{\beta_4hash} \times e^{\beta_5category_1} \times \dots \times e{\beta_jcategory_{j+4}}
\end{equation}

\section{Results}
\subsection{The Social Sciences are associated with more Offensive Tweets}
We found 12,380 out of 760,799 original tweets to be offensive (1.63\%, see Table \ref{table1}). Note that a tweet can be both offensive and negative, as they are results from two different classifiers (one for valence, another for offensive language).

We examine the relationship between academic disciplines and the rates of offensive classifications exhibited in each discipline. As shown in Table \ref{table2}, there is a clear distinction between the rates of offensive language in different disciplines. We test their association with a Kruskal-Wallis rank sum test, as the underlying distributions are dissimilar.  The results determine that academic discipline is strongly associated with offensive language ($X^2=5383, p < 2.2e-16$).

Tweets made about Philosophy papers have the highest rates of offensive language, with 5.2\% of all unique tweets in Philosophy being classified as offensive. Furthermore, we find that the Humanities and the Social Sciences have the highest rates of offensive classifications; out of the eight disciplines with the highest incidents of tweets with offensive language, six (Philosophy, Political Science, Psychology, Sociology, Geography, Economics) are either Humanities or Social Sciences. The two disciplines with the least incidents of offensive language are Math (.9$\%$) and Engineering (.9$\%$). We can relate this back to the politicization of science, and the implied connection between politicization and offensive language. Humanities and Social Sciences are political by nature, and hence there is an intuitive justification for their strong representation in Table \ref{table2}. Furthermore, the two sciences that are represented in the top 2 (Geology, 3.3$\%$, Environmental Science, 2.3$\%$) are concerned with subjects of controversy such as global warming and fossil fuels.

\begin{table}[]
\caption{Classification results}
\centering
\begin{tabular}{@{}llll@{}}
\toprule
 & Total Tweets & \%  \\ \midrule
Negative & 153,860 & 20.22 \\
Neutral & 490,783 & 64.51 \\
Positive & 116,156 & 15.27 \\
Offensive & 12,380 & 1.63 \\ \midrule
Overall & 760,799 & \\ \bottomrule
\end{tabular}
\label{table1}
\end{table}

\begin{table}[]
\caption{Rates of offensive language by discipline, sorted.}
\begin{tabular}{@{}llll@{}}
\toprule
 & Total Tweets & Offensive & \%  \\ \midrule
Philosophy & 10,714 & 555 & 5.2 \\
Geology & 32,399 & 1,081 & 3.3 \\
Political Science & 91,742 & 2,271 & 2.5 \\
Psychology & 114,975 & 3,671 & 2.5 \\
Environmental Science & 53,897 & 1,236 & 2.3 \\
Geography & 62,616 & 1441 & 2.3 \\
Sociology & 44,510 & 946 & 2.1 \\
Economics & 59,725 & 1,137 & 1.9 \\
Materials Science & 9,646 & 142 & 1.5 \\
Biology & 228,964 & 3540 & 1.5 \\
History & 10,557 & 152 & 1.4 \\
Chemistry & 31,031 & 423 & 1.4 \\
Business & 45,610 & 602 & 1.3 \\
Computer Science & 121,567 & 1534 & 1.3 \\
Art & 4,604 & 58 & 1.3 \\
Medicine & 415,233 & 5,354 & 1.3 \\
Physics & 36,113 & 390 & 1.1 \\
Math & 27,378 & 259 & 0.9 \\
Engineering & 28,148 & 266 & 0.9 \\
 \bottomrule
\end{tabular}
\label{table2}
\end{table}

\begin{table}[]
\caption{Sample of topics for each topic category. N denotes the number of topics in the topic category.}
\resizebox{\textwidth}{!}{
\begin{tabular}{|c|c|}
\hline
\multirow{4}{*}{Science(N=24)}
& gene eye decreases gmo glyphosate regions expression midbrain \\& fore sucks throwing regulate arse kale glucose \\\cline{2-2}
& denialist address evidence cited willfully sources scientific \\& paranoid lack free blogs debunks troll disingenuous needed \\\hline
\multirow{4}{*}{Political(N=11)}
& nuclear noncombatants hiroshima revisiting weapons iran makes \\& dumb smartphone dope phone dumber smart politics bonkers\\\cline{2-2} 
& sold sales unethical tactics bullshit clinton salesman manipulator \\& scale notmypresident logic statements blow lacks term                 \\\hline
\multirow{4}{*}{Race(N=2)}
& biased holy crap narrative false horribly racial \\& presents adult translation homicides article mit cking ethnic\\ \cline{2-2} 
&black white racist racism pain thicker anxiety \\& skin ppl doctors cops voted worry heroin opioids
\\\hline
\multirow{4}{*}{Gender and Sexuality(N=7)} 
& men volume active women artery sexually coronary monogamous plaque \\& testosterone relationships treatment greater patriarchy bitch \\\cline{2-2} 
& women gay heterosexual lesbian experiences orgasm ages \\& sample pleasure touching frequency genital bisexual sexual differences  \\ 
\hline
\multirow{4}{*}{Religion(N=3)}
& kitchen sponges vertebrate kinds billions sterilizing humans \\& islam united monster racialization bacteria islamophobia flying brown \\\cline{2-2} & brain damage religious sleep injury chicken fundamentalism eats \\& suffering head altered function dysbiosis pox woodpeckers    \\ 
\hline
\multirow{4}{*}{Other(N=3)}            & fucked dick fascinating model title lot enjoy \\& poop yep worth travel neanderthal explain jet deranged \\\cline{2-2} & holy shit cow sherlock crock bull wild umm coolest \\& smh moment late potentially mirror publishes                                                            \\
\hline
\end{tabular}
}
\label{table4}
\end{table}

\begin{table}
\caption{Most frequent topic and topic category per discipline. 'n' denotes the occurrences of the most frequent topic and topic categories followed by percentage of the most common topic and topic category for each discipline.}
\centering
\small
\resizebox{\columnwidth}{!}{%
\begin{tabular}{|>{\hspace{0pt}}m{0.19\linewidth}|>{\hspace{0pt}}m{0.59\linewidth}|>{\hspace{0pt}}m{0.16\linewidth}|} 
\hline
                                     & Topic                                                                                                                                                       & Category                        \\ 
\hline
Biology              & guns killing tory regime austerity disabled british electorate tories jews sick government nazis chemo vulnerable  (n=212, 10\%)                                          & Science (n=998, 48\%)  \\ 
\hline
Business             & guns killing tory regime austerity disabled british electorate tories jews sick government nazis chemo vulnerable (n=28, 10\%)  & Science (n=114, 41\%)             \\ 
\hline
Chemistry            & ignorance useless willful supplement predict bitches neurotoxicity adjuvant alum save bcaa rubbish tests aluminum deceit (n=15, 10\%)                                    & Science (n=93, 60\%)                        \\ 
\hline
Computer Science     & holy shit cow sherlock crock bull wild umm coolest smh moment late potentially mirror publishes  (n=28, 08\%)                                                           & Science (n=176, 53\%)                          \\ 
\hline
Economics            & nuclear noncombatants hiroshima revisiting weapons iran makes dumb smartphone dope phone dumber smart politics bonkers (n=10, 8\%)                                     & Science (n=54, 43\%)                      \\ 
\hline
Engineering           & associated profound views misperceiving cruz rubio bullshit trump conservatism favorable plosone plos drumpf assoc comments (n=2, 15\%)                                & Science (n=8, 62\%)                      \\ 
\hline
Environmental Science & denialist address evidence cited willfully sources scientific paranoid lack free blogs debunks troll disingenuous needed (n=98, 10\%)                                   & Science (n=609, 62\%)                        \\ 
\hline
Geography            & giant dinosaurs australia pee hole mass frog massive million banned frogs sun birds pandemic close (n=22, 6\%)                                                        & Science (n=211, 57\%)                        \\ 
\hline
Geology               & satellite hockey stick mann dishonest data denialist anomaly moth question michael multiple pretends blonde questions  (n=25, 8\%)                                     & Science (n=191, 57\%)                          \\ 
\hline
History                & denialist address evidence cited willfully sources scientific paranoid lack free blogs debunks troll disingenuous needed (n=3, 12\%)                                  & Science  (n=12, 48\%)                       \\ 
\hline
Materials Science    & fuck fucking cool leopard chimpanzees kidding yeah crispr metal cas urine bitch cephalopods chimps milk (n=10, 19\%)                                                    & Science                         (n=40, 74\%) \\ 
\hline
Mathematics           & reviewed peer published illness paper journal suspects opinion garbage alcohol mental dysbiosis nobel hippopotamus domestic (n=2, 18\%)                                  & Science (n=6, 55\%)                        \\ 
\hline
Medicine             & guns killing tory regime austerity disabled british electorate tories jews sick government nazis chemo vulnerable  (n=320, 83\%)                                          & Science (n =1544, 50\%)              \\ 
\hline
Philosophy          & guns killing tory regime austerity disabled british electorate tories jews sick government nazis chemo vulnerable (n=148, 42\%)                                          & Political (n=238, 68\%)             \\ 
\hline
Physics              & fuck fucking cool leopard chimpanzees kidding yeah crispr metal cas urine bitch cephalopods chimps milk (n=13, 12\%)                                                  & Science (n=68, 63\%)                        \\ 
\hline
Political Science     & russia you re hypocrite elections smells america uranus baloney sponsored election farts elected infanticide cia (n=54, 8\%)                                           & Political (n=292, 45\%)                      \\ 
\hline
Psychology          & gender fractions feminist fungus lipid glaciology ants glaciers looks sounds surgeon like sound takes badass (n=92, 4\%)                                               & Science (n=995, 42\%)  \\ 
\hline
Sociology            & gender fractions feminist fungus lipid glaciology ants glaciers looks sounds surgeon like sound takes badass (n=17, 8\%)                                                & Science (n=98, 44\%)  \\
\hline
\end{tabular}
}
\label{table5}
\end{table}

\subsection{Topic Categories describe a wide variety of Controversial Topics}
We find that $K$=50 topics had the highest average coherence score among its topics ($c_v=.285$). Furthermore, the model was able to maintain a sufficiently high level of topic uniqueness, with a topic uniqueness of 1 implying complete uniqueness between topics. Maintaining a high level of topic uniqueness allows us to obtain more discrete decompositions of offensive tweets into topics for the regression analysis, and further allows us to obtain more clear distinctions among topics.

\subsubsection{Summary Statistics}
Table \ref{table4} shows two topics from each topic category, and the number of topics $N$ contained in the topic category. Table \ref{table5} shows the most frequent topic and topic category for each academic discipline. We also display the number of tweets $n$ and percentage of tweets within that discipline associated with their most frequent topic and topic category.

\subsubsection{Science}
We determined that 24 of our topics were pertained strongly to discussions of science. Extracting the most representative tweets for the first topic in Table \ref{table4} under the category of science reveals pointed discussions of GMOs, herbicides, mental health, diabetes, gene modification, and global warming. These are visibly related to the words ``gene'', ``gmo'', ``midbrain'', ``glucose'', and ``regulate'' found within the topic. The second topic, which is also the most represented topic for offensive tweets that reference an environmental science DOI, is clearly related to denialism of global warming upon digging into its most representative tweets. The most representative offensive tweet of its topic cites an academic article on recent mass extinction in oceans in an argument with another user: ``Yes, you are a mindless troll. Not my fault you still lack the intellect + integrity to address scientific research.'' This supports previous literature suggesting that toxic language arises in political arguments on social media \cite{mosleh-ea-2021, anderson-huntington-2017}.

Looking at other scientific disciplines and their most frequent topics also show clear relationships between mentions of controversial topics and how offensive language occurs within these scientific topics. For example, tweets related to a topic that mentions ``crispr'' both express wary and excitement for it by usage of profane language. Analysis of Geology's most frequent topic gives further proof of how offensive language is related to politicization of science; the abstruse term ``hockey stick'' in the topic is shown to refer to the ''hockey stick graph'' of temperature anomaly \cite{marsicek-ea-2018}. CDA also lets us analyze the broader conversations that these tweets occur in, and the tweets that do not reference the original DOI that are found in the responses. In response to one offensive tweet that mentions the ``hockey stick,'' a user writes ``Seriously, what are you talking about. I can’t believe you call yourself a climate scientist... The evidence for these warm periods is everywhere, none of this is even disputed. Sigh.'' The nature of offensive language in scientific topics seem to occur frequently in debates on both sides of a polarizing subject, and as well as highlighting criticism or support for new scientific discoveries.

An important takeaway from this category is that we can readily reject an intuition that citing academic articles with aggressive language is primarily used for ``debunking'' false information. What we find instead is a mixture of both; aggressive language positively highlighted climate change articles but were also used in tweets such as the ``hocky stick'' tweet, where an article or an excerpt from an article is ``cherry-picked'' to maintain science-denialism. 

\subsubsection{Race}
The two topics pertaining to race in our topic model are represented by discussions of racialized violence. The first topic implies connections between race, ethnicity, and violence (``homicide''); the second is an amalgamation of racialized violence through police and the opioid epidemic. An analysis of the most representative tweets reveals angered discussions on police brutality (``cops are racist shits''), domestic violence, and Trump's racist policies and language. While many of the tweets reviewed act in the spirit of anti-racist activism, there are several tweets that conflate race with intelligence and crime.

The emergence of actively racist tweets in this sample best indicate how academic articles are reshaped and reimagined to fit existing ideologies from multiple viewpoints. The dialectic between anti-racism and racist tweets highlights that the inflammatory usage of academic articles is not limited to simply one perspective, and that the the weaponization of academic articles is amorphous; similar to the category of science, we find that multiple perspectives exist, and that the articles are both used to push back and to maintain false and dangerous ideologies.
\subsubsection{Religion}
Topics related to religion reference islamophobia, terrorism, and religious fundamentalism. Seemingly random discursive words are embedded into these topics, such as ``kitchen sponges'' in the first, and ``brain damage'' in the second.  In the most representative sample of tweets belonging to this category of topics, we find that these tweets only contain tweets that maintain the singular attitude of being islamophobic, often conflating Islam with terrorism. The question of how ``brain'' is incorporated into the second topic is revealed through looking at the most representative tweets; several tweets perform a selective reading on an article to conclude that religious fundamentalism is related to brain damage. This category is thus notable for its homogenous perspectives, which depart from other categories that reveal different attitudes and ideologies. 

\subsubsection{Politics}
Political topics made up the second largest category of topics. We gave thirteen topics a primary category of politics. Any topic that contained words that signified a branch or segment of the government, or the name of a specific politician, was considered for this topic category. Several of the tweets that are most representative of the first topic contain tweets that refer to nuclear weapons, marijuana legalization, nuclear energy, and nuclear energy. These tweets are heavily polarized towards the ends of the political axis; a tweet that mentions identity politics uses “dumb” to describe identity politics as a “manifestation of the left.” Tweets that mention nuclear energy link DOIs related to the efficiency of nuclear energy to push back on the conservatives’ propensity towards fossil fuels. In one tweet, its user uses profane language to argue that conservative voters’ tendency to believe misinformation.

Offensive tweets within this category take on both conservative and liberal partisan slants. We categorized the topic that has the most tweets that belong to it within the disciplines of biology, business, medicine, and philosophy as political; this topic, which contains terms such as “guns,” “killing,” “tory,” “electorate,” “government,” reveals several different political issues after examining the tweets that represent it. Tweets discuss war, abortion, wealth inequality, healthcare, and gun violence. The nature of several of these tweets is anticipatory of political backlash from conservative spheres; in one tweet, a user writes that an article’s findings will be contested as being written by “snowflakes” or “nazis.” We see the language that is commonly viewed as being used in conservative spheres being reappropriated to anticipate hostility; another tweet that is assigned this topic suggested that wealth inequality is not just a “Leftist crank issue.” Our qualitative analysis suggests that several of these tweets use offensive words in conjunction with conservative “dog whistles,” which are words that are contextually understood within a specific political context, such as “snowflake” or “libtard.” However, the political undercurrent of the tweets in our corpus suggest a left-leaning political stance, and use these terms not to prove citizenship in conservative networks on Twitter, but rather in anticipation of political backlash. This can be seen as a measure that shields against ad-hominem claims against academics, which has been shown to lessen the validity of claims made by scientists \cite{barnes-2018}. Of the 733 tweets that are assigned this topic category, only a handful expressed concrete conservative sentiments, while most expressed left-leaning sentiments.

\subsubsection{Gender and Sexuality}

We generated the topic category of Gender and Sexuality from seven different topics. The nature of the topics and tweets that have their strongest alignment with this topic category are heavily critical of misogyny, sexual violence against women, constrictive abortion laws, sexuality-based prejudices, and sexual freedom. Similar to the tweets that are associated with the political topic category, these tweets used aggressive and offensive language defending feminist practices and studies. Conversely, multiple tweets criticize findings on womens’ sexual practices along a conservative slant.
\subsubsection{Other}

We generated the topic category of “other” from extraneous topics and their associated tweets that did not have an easily visible theme. Tweets that had the highest likelihood of being assigned a topic within these three topics had no clear grouping. However, qualitative analysis of this category’s topics and associated tweets reveal significantly high levels of profanity. Their usage of profanity differs; for example, tweets associated with the first topic use directed offensive language, to borrow from Waseem et al. \cite{waseem-etal}’s typology of abusive language. The second topic, in particular, contains multiple tweets that express “holy shit” to various academic articles, which are a generalized form of profanity.

\subsection{Offensive Tweets are Retweeted Less, but Faster}
\subsubsection{Retweet Volume}

We perform a negative binomial regression analysis to answer whether or not offensive language in tweets that reference academic articles increase or decrease their virality. Results indicate that negative sentiment  ($\beta$=.045, p \textless{} 2e-16), number of followers ($\beta$=1.05, p \textless{} 2e-16), presence of a hashtag ($\beta$=.14, p \textless{} 2e-16), and offensive language ($\beta$=-.25, p\textless{}2e-16) are significant factors for determining the total volume of retweets for a tweet in our dataset. Our analysis shows that log(followers) and presence of a hashtag are the strongest factors for increasing retweet count. Offensive language, on the other hand, is expected to decrease the retweet count. The strength of a regression coefficient $\beta$ can be evaluated with $e^\beta$. An offensive tweet is expected to reduce the number of retweets by $1 - e^{-.2538}$, or 22.4\%, compared to non-offensive tweets. Negative sentiment, while significant, only affects retweet volume very marginally ($e^\beta=1.046$), and positive sentiment is insignificant.

\begin{table}[]
\caption{Regression results for $numRT$ \\ *** significant at .1 percent; ** significant at 1 percent; * significant at 5 percent }
\begin{tabular}{llll}
\toprule
\multicolumn{1}{l}{$numRT$} &  &  &  \\ \midrule
 & $\beta$ & std. error & Pr(\textless{$|$z$|$}) \\

hash*** & 0.144315 & 0.004751 & \textless{}2e-16 \\

positive & 0.010191 & 0.005651 & 0.0713 \\

negative*** & 0.043042 & 0.006454 & \textless{}2.58e-11 \\

log(followers)*** & 1.047963 & 0.008192 & \textless{}2e-16 \\

Political* & -0.095680 & 0.044638 & 0.0321\\

Race & -0.148262 & 0.100263 & 0.1392\\

Religious & 0.157492 & 0.085436 & 0.0653\\

Other*** & -0.361750 & 0.074653 & \textless{}7.54e-10\\

Gender and Sexuality*** & -0.329844 & 0.068455 & \textless{}1.45e-6\\

Science*** & -0.384225 & 0.045510 & \textless{}2e-16\\
 &  &  &  \\ \midrule
${R}^2$ & .567 &  \\
N. Observations & 295,078 
 \\ \bottomrule
\end{tabular}%
\label{table7}
\end{table}

Table \ref{table7} shows the results for our regression analysis when we decompose the offensive language factor into topic categories, as a tweet belonging to a topic category necessarily indicates that it is classified as offensive. We find nearly equivalent regression coefficients for the unchanged factors. We identify that offensive tweets in the topic categories of political ($\beta$=-.09, p = .032), other ($\beta$=-.36, p \textless{} .754e-10), gender and sexuality ($\beta$=-.33, p \textless{} 1.45e-6), and science ($\beta$=-.38, p \textless{} 2e-16) are statistically significant, and decrease the number of retweets a tweet receives. Offensive tweets associated with the topic categories of race were not deemed significantly significant.

\subsubsection{Retweet Speed}

To test our independent factors against regression speed, we filter our dataset for tweets with at least 25 retweets (N=6949). We should note that the number of offensive tweets in this sample is very small (n=54).

Regression analysis for the difference in minutes between the original tweet and 25th retweet that hashtags ($\beta$=.35, p \textless{}8.52e-10), negative sentiment ($\beta$=.18, p=0.0038), and positive sentiment ($\beta$=.15, p=.02) increases the time between the original and 25th retweet, and hence decreases retweet speed. We find that offensive language ($\beta$=-2.06, p \textless{}3.46e-10) and log(followers) ($\beta$=-0.886, p \textless{}2e-16) substantially increase retweet speed, by nearly 87\% and 59\%, respectively.

\section{Discussion}

Our analysis of offensive language in tweets that reference academic articles confirms intuitions about entanglement of offensive language and the politicization of science. Our CDA and topic modeling confirm several outstanding findings from previous literature. Firstly, they bolster previous research that finds offensive language within politicized contexts on Twitter \cite{anderson-huntington-2017, mosleh-ea-2021, kong-ea-2020}. Secondly, they reveal incredibly varied subjects of political contention such as but not limited to global warming, racialized violence, abortion, and sexuality. We further emphasize that topic categorizations are not completely discrete; by the nature of topic modelling, certain topics can contain different themes that can be split between the topic categories we interpreted. However, this inability for our generated topics to be categorically distinct within our topic categories reflects the shared political undercurrent carried in each of our topic categories. Outside of our topic category of ``Other,'' our topic categories of Race, Science, Politics, Gender and Sexuality cannot be disentangled from politics.

Specifically, our results indicate that offensive language is weaponized at a significant rate across many disciplines, and for a wide variety of uses. This reveals that science dissemination, at the fringes, may have dangerous characteristics that have not been identified before. Firstly, we see that this type of dissemination of science carries strains of racism, science denialism, homophobia, gender discrimination, and islamophobia that reinterprets and selectively cites academic articles. The fact that they are used for argumentation should be a concern; these offensive tweets reach a level of virality quicker than non-offensive tweets (though they reach a narrower audience). This calls for more investigations on how science is being cited in discursive arenas on social media as views and credibility of science are changing.

Though diffusion of academic articles is vital for researchers' success, academics should not be attacked with offensive language or have their works weaponized in incivil attacks, as our CDA has shown. Academia is not a context where the axiom ``any publicity is good publicity'' applies, and especially when its publicity is in conjunction with abusive and offensive language; research has found that adhominem attacks and the use of abusive language in scentific discussions leads to lower perceived credibility \cite{barnes-2018, konig-2019}. Favorably, our results indicate that offensive tweets referencing an academic DOI reach a smaller audience than non-offensive tweets. However, offensive tweets in our dataset that do reach a level of virality do so much quicker than non-offensive tweets.

\section{Conclusion}
This work conducts mixed-method analyses of offensive tweets that references academic articles, and the relationship between offensive tweets, politicization of science, and virality. We use sentiment and offensive language classification to annotate our dataset of tweets, and then perform critical discourse analysis over our generated topic models to contextualize our topics with the broader theme of the politicization of science. Our critical discourse analysis reveals that offensive language is heavily utilized in politicized messages regarding topics such as race, science, gender and sexuality, religion. Lastly, results from our regression analysis show that offensive tweets that reference academic DOIs are diffused at lower volumes, but have rapid diffusion when they do go viral.
\subsection{Limitations}
We should note that our dataset contains only tweets from the 9,650 most mentioned articles on Altmetric, with the most recent tweet in the dataset stemming from nearly 4 years ago. This presents several limitations. Firstly, selecting tweets mentioning only popular articles may imply higher quality articles from more reputable authors and publications, which may imply lower inherent frequencies of offensive language in the tweets that reference them. Secondly, the current political atmosphere of ``X'', formerly known as Twitter, may not be well represented in our study. Several large events have occurred that may drastically change the behaviors of offensive language on Twitter and are not captured in our dataset, such as the COVID-19 pandemic, resurgence of BlackLivesMatter during George Floyd protests, and the 2020 American election. Though our dataset does not contain these events, they have clear relationships with our work's findings. Lastly, Twitter has changed hands, as Elon Musk took over Twitter in October 2022, and has rebranded it to ``X'' as well as changed several of the platform's rules and guidelines. 

\begin{acks}
We would like to acknowledge for their continued support and many reviews of this paper from its infancy to where it is now. 
\end{acks}
\bibliographystyle{ACM-Reference-Format}
\bibliography{sample-base}

\end{document}